# Bifurcations and Chaos in Simple Dynamical Systems

**Mrs. T.Theivasanthi, M.Sc., M.Phil.**
Lecturer in Physics, PACR Polytechnic College, Rajapalayam – 626117, India
Email: sankarg4@yahoo.com

## Abstract

Chaos is an active research subject in the fields of science in recent years. It is a complex and an erratic behavior that is possible in very simple systems. In the present day, the chaotic behavior can be observed in experiments. Many studies have been made in chaotic dynamics during the past three decades and many simple chaotic systems have been discovered. In this work, **"Bifurcations and Chaos in Simple Dynamical Systems"** - the behavior of some simple dynamical systems is studied by constructing mathematical models. Investigations are made on the periodic orbits for continuous maps and idea of sensitive dependence on initial conditions, which is the hallmark of chaos, is obtained.

## Lead Paragraph

A SMALL ATTEMPT HAS BEEN MADE TO FIND OUT THE REASONS / UNKNOWN CONDITIONS FOR THE PRODUCTION OF CHAOS IN A SYSTEM. THIS IS EXPLAINED THROUGH SIMPLE DYNAMICAL SYSTEMS. (WHY CHAOS IS PRODUCED IN A FORCED DAMPED SIMPLE PENDULUM?) BESIDES, ANOTHER ATTEMPT HAS BEEN MADE TO IDENTIFY, ALGEBRAICALLY SIMPLEST CHAOTIC FLOW. THESE ARE THE SIGNIFICANCE OF THIS STUDY - "BIFURCATIONS AND CHAOS IN SIMPLE DYNAMICAL SYSTEMS". ACCORDINGLY, AN ANALYSIS IS DONE ON DIFFERENT DYNAMICAL SYSTEMS. THE EXACT SOLUTION IS OBTAINED BY SOLVING THE DIFFERENTIAL EQUATION BY USING RUNGE-KUTTA METHOD – AS A RESULT, IT IS CLEAR FROM THE ANALYSIS THAT, *PERIOD MULTIPLICATION* OCCURRING IN A FORCED DAMPED SIMPLE PENDULUM WHICH LEADS TO CHAOS. THIS RESULT IS PROVING THAT IT IS POSSIBLE TO FIND OUT THE REASONS / UNKNOWN CONDITIONS UNDER WHICH CHAOTIC BEHAVIOUR EXHIBITS IN VARIOUS SYSTEMS. THE IMPORTANCE OF RESULT IS, "WHY CHAOS IS PRODUCED IN VARIOUS SYSTEMS? - MAY BE IDENTIFIED IN FUTURE.

# Chapter – I  Introduction

Chaos has been a subject of active research in the fields of physics, mathematics and in many other fields of science in recent years. It is an erratic behavior that is possible in very simple systems. In the present day, scientists realize that the chaotic behavior can be observed in experiments and computer models of behavior from all fields of science. It is now common for experiments, whose previous anomalous behavior was attributed to experiment error or noise, to be reevaluated for an explanation in these new terms.

During the past three decades, extensive studies have been made in chaotic dynamics [1 – 5]. A Dynamical system consists of a set of possible states together with a rule that determines the present state in terms of past states. The rule ***deterministic*** is, if we can determine the present state uniquely from the past states. But, if there is a randomness in our rule that is called a random or stochastic process, for example, a mathematical model for the price of gold as a function of time would be to predict today's price to be yesterday's price plus or minus one dollar with the two possibilities equally likely.

If the rule is applied at discrete times, it is called a ***discrete – time dynamical system*** which is also called as ***maps***. A discrete – time dynamical system takes the current state as input and ***updates*** the situation by producing a new state as output.

The other type of dynamical system is the limit of discrete system with smaller and smaller updating times. The governing rule in that case becomes a set of differential equations and the term ***continuous – time dynamical system*** is used.

Since the seminal work of Lorenz in 1963 and Rossler in 1976 [5], it has been known that complex behaviour i.e. Chaos can occur in systems of autonomous Ordinary Differential Equations (ODES) with a few as three variables and one or two quadratic nonlinearities. Many other simple chaotic systems have been discovered and studied over the years. With the growing availability of powerful computers, many other examples of chaos have been subsequently discovered. Yet the sufficient conditions for chaos in a system remain unknown.

Using extensive computer search, J.C.Sprott [4], (J.C.Sprott, Dept.of Physics, University of Wisconin-Madison) found 19 distinct chaotic models with three dimensional vector fields that consist of five terms including two non-linearities and six terms with one quadratic non-linearity.

Vinod Patidar and K.K.Sud [7], investigated the global dynamics of a special family of jerk systems which has a non-linear function and they are known to exhibit chaotic behaviour at some parameter values. They [8], have also made a through investigation of synchronization of identical chaotic jerk dynamical systems. Since, from practical point of view, one would like to convert chaotic solutions into periodic limit cycle or fixed point solutions.

In this present work, the behavior of some simple dynamical systems is studied by constructing mathematical models. Investigations are made on the periodic orbits for continuous maps and idea of sensitive dependence on initial conditions, which is the hall mark of chaos, is obtained.

The first chapter gives an introduction on different types of dynamical systems. The second chapter explains fixed points, cobweb plot and stability of fixed points. The logistic model is studied in the third chapter where the concept of non-linearity is introduced. The family of logistic map is investigated for different parameter values. Bifurcation diagram is drawn to show the birth, evolution and death of attracting sets. The impact of sensitive dependence on the initial measurements on the orbit in a two dimensional map is worked out in the fourth chapter. In addition to this, two physical processes are modeled with maps and the use of maps in scientific applications is discussed.

# Chapter – 2  Logistic Model

We study models because they suggest how real – world processes behave. Every model of a physical process is at best an idealization. The goal of a model is to capture some feature of the physical process. The feature we want to capture now is the patterns of the points on an orbit. In particular, we will find that the patterns are sometimes simple and sometimes quite complicated or chaotic even for simple maps.

The function *f(x)= 2x* is a simple mathematical model, where x denotes the population of bacteria in a laboratory culture and *f(x)* denotes the population one hour later. If the culture has an initial population of 10,000 bacteria then after one hour, there will be *f(10,000) = 20,000* bacteria and after two hours *f[f(10,000)] = 40,000* bacteria and so on.

But, this growth cannot continue for ever. At some point the resources of the environment will become compromised by the increased population and the growth will slow to something less than exponential.

An improved model to be used for a resource limited population might be given by *g(x)=2x(1-x)*. This is a non linear effect and the model is an example of logistic growth model.

**Table – 1   Comparison Of Exponential Model And Logistic Model**

| n | f(x)=2x | g(x)=2x(1-x) |
|---|---|---|
| 0 | 0.01 | 0.01 |
| 1 | 0.02 | 0.0198 |
| 2 | 0.04 | 0.038816 |
| 3 | 0.08 | 0.074618 |
| 4 | 0.16 | 0.138101 |
| 5 | 0.32 | 0.238058 |
| 6 | 0.64 | 0.362773 |
| 7 | 1.28 | 0.462338 |
| 8 | 2.56 | 0.497163 |
| 9 | 5.12 | 0.499984 |
| 10 | 10.24 | 0.5 |
| 11 | 20.48 | 0.5 |
| 12 | 40.96 | 0.5 |

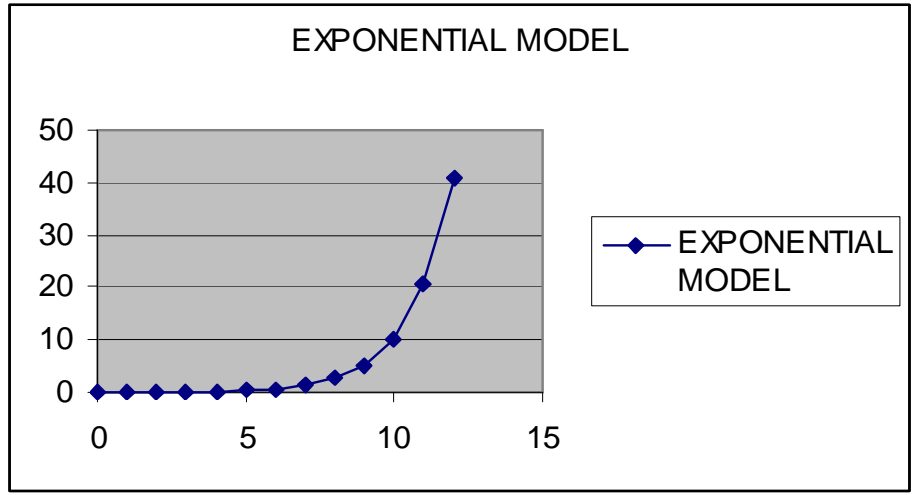

**Fig.1**

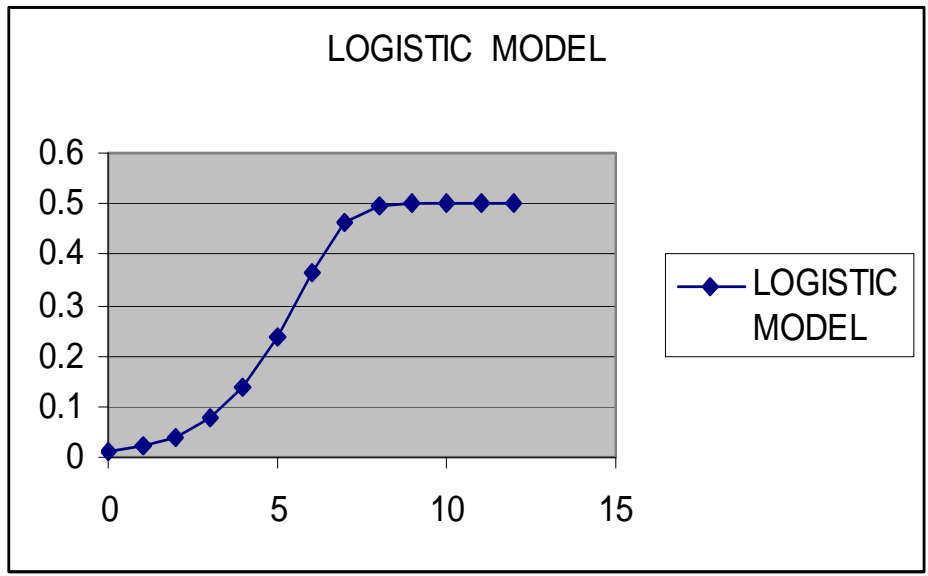

**Fig.2**

From the table [1], it is evident that for the function $g(x)=2x(1-x)$ the population approaches an eventual limiting size which we would call a steady state population. The population saturates at $x=0.5$ and then never changes again. The exponential model explodes while a logistic model approaches a steady state.

If we use starting population other than $x=0.01$, the same limiting population $x=0.5$ will be achieved.

Table – 2  **Showing Starting Populations Between 0.0 And 1.0**

| g(x) for STARTING POPULATION X = | | | | | | | | |
|---|---|---|---|---|---|---|---|---|
| 0.1 | 0.2 | 0.3 | 0.4 | 0.5 | 0.6 | 0.7 | 0.8 | 0.9 |
| 0.18 | 0.32 | 0.42 | 0.48 | 0.5 | 0.48 | 0.42 | 0.32 | 0.18 |
| 0.2952 | 0.4352 | 0.4872 | 0.4992 | 0.5 | 0.4992 | 0.4872 | 0.4352 | 0.2952 |
| 0.416114 | 0.491602 | 0.499672 | 0.499999 | 0.5 | 0.499999 | 0.499672 | 0.491602 | 0.416114 |
| 0.485926 | 0.499859 | 0.5 | 0.5 | 0.5 | 0.5 | 0.5 | 0.499859 | 0.485926 |
| 0.499604 | 0.5 | 0.5 | 0.5 | 0.5 | 0.5 | 0.5 | 0.5 | 0.499604 |
| 0.5 | 0.5 | 0.5 | 0.5 | 0.5 | 0.5 | 0.5 | 0.5 | 0.5 |
| 0.5 | 0.5 | 0.5 | 0.5 | 0.5 | 0.5 | 0.5 | 0.5 | 0.5 |
| 0.5 | 0.5 | 0.5 | 0.5 | 0.5 | 0.5 | 0.5 | 0.5 | 0.5 |
| 0.5 | 0.5 | 0.5 | 0.5 | 0.5 | 0.5 | 0.5 | 0.5 | 0.5 |
| 0.5 | 0.5 | 0.5 | 0.5 | 0.5 | 0.5 | 0.5 | 0.5 | 0.5 |
| 0.5 | 0.5 | 0.5 | 0.5 | 0.5 | 0.5 | 0.5 | 0.5 | 0.5 |

For this logistic model $X=0.5$ is the ***fixed point***

A function whose input and output are the same will be called a map. A point $p$ is a fixed point of the map $f$ if $f(p)=p$ for e.g for $g(x)=2x(1-x)$, the fixed point are $x=0$ and $x=1/2$.

### Sink

If all points sufficiently close to $p$ are attributed to $p$, then the $p$ is called a sink or an attracting fixed point.

### Source

If all points sufficiently close to *p* are repelled from *p*, then the *p* is called a source or a repelling fixed point.

### Smooth Function

A type of function for which the derivatives of all orders exist and are continuous is called a smooth function. If $|f^1(p)| < 1$, then *p* is a sink. If $|f^1(p)| > 1$ then *p* is a source.

### 2.1. COBWEB PLOT

A cobweb plot illustrates convergence to an attracting fixed point of $g(x)=2x(1-x)$. Let $x_o=0.1$ be the initial condition. Then the first iterate is $x_1=g(x_o)=0.18$.

Note that the point $(x_0,x_1)$ lies on the function graph and $(x_1,x_1)$ lies on the diagonal line. Connect these points with a horizontal dotted line to make a path. Then $x_2=g(x_1)=0.2952$ and continue path with a vertical dotted line to $(x_1,x_2)$ and with a horizontal dotted line to $(x_2,x_2)$. An entire orbit can be mapped out this way. The orbit will converge to the intersection of the curve and the diagonal $x=1/2$.

If the graph is above the diagonal line y=x, the orbit will move to the right; if the graph below the line, the orbit will move to left.

.

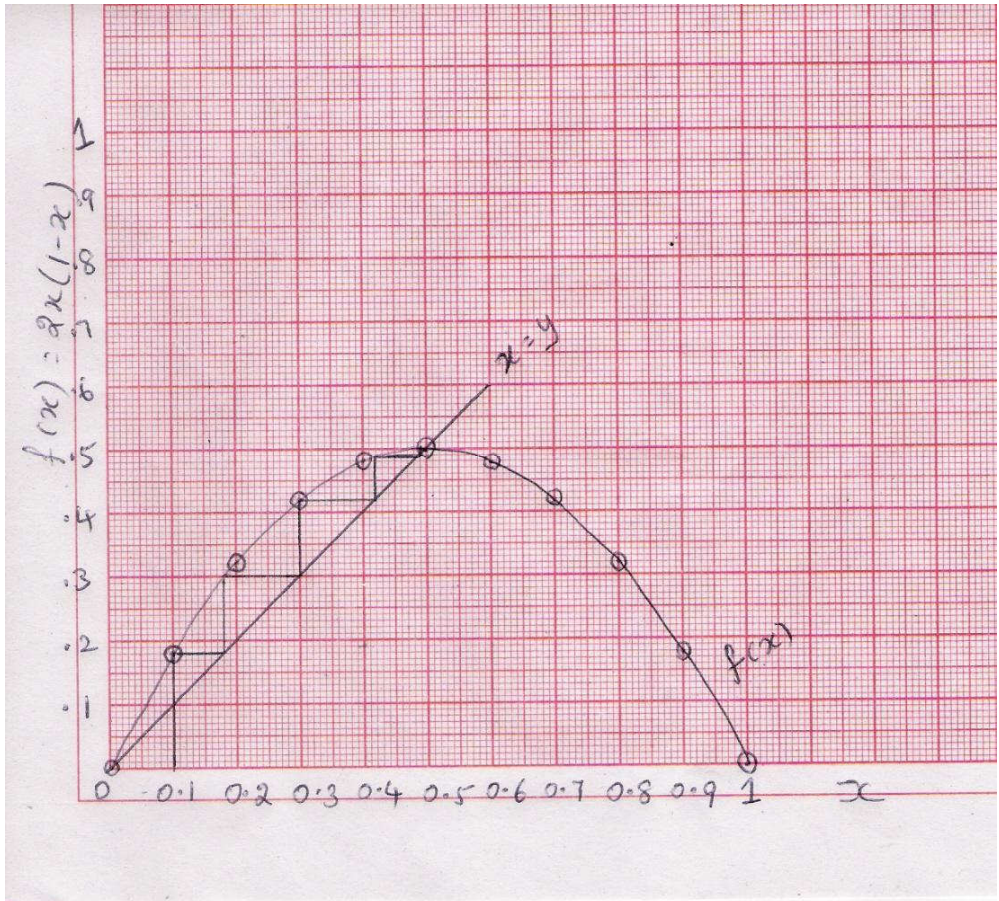

**Fig.3   Cobweb Plot   *f(x)=2x(1-x)***

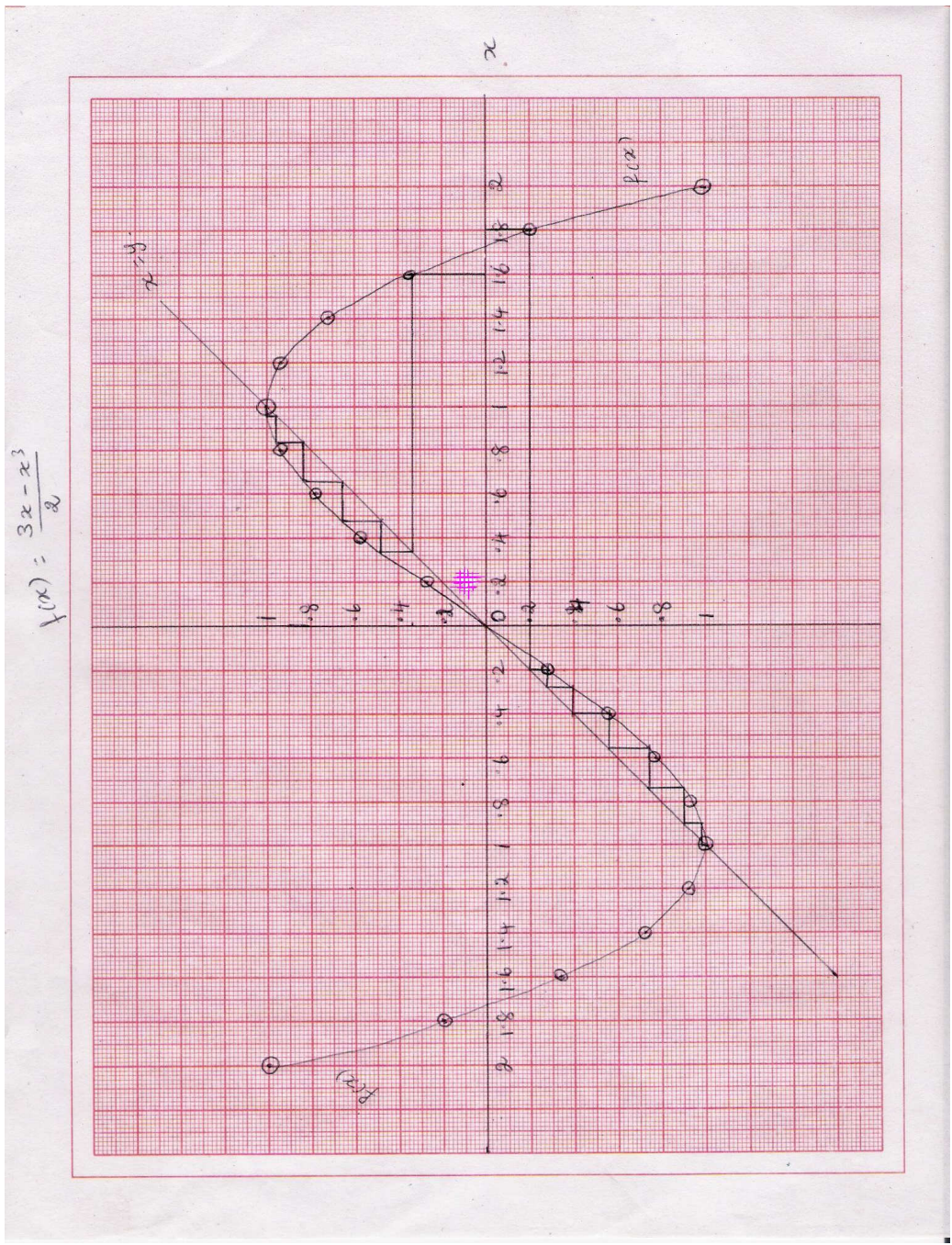

**Fig.4** <u>Cobweb Plot</u>   $f(x)=(3x-x^3)/2$

## 2.2. Stabilty Of Fixed Points

A stable fixed point has the property that points near it are moved even closes to the fixed point under the dynamical system. For an unstable fixed point, nearby points move away as time progresses.

The question of stability is significant because a real world system is constantly subject to small perturbations. Therefore, a steady state observed in a realistic system must correspond to a stable fixed point. If the fixed point is unstable small errors or perturbations in the state would cause the orbit to move away from the fixed point.

## Results and Discussion

For the map $f(x) = 2x(1-x)$, the fixed points are found by solving the equation $x = 2x(1-x)$. There are two solutions, $x = 0$ and $x = 0.5$ which are two fixed points of $f(x)$. The $x = y$ line cuts the function graph at the fixed points. The orbit of the function with initial value 0.1 is drawn. The orbit converges to the sink at $x = 0.5$

The fixed points are found in the same manner for the map $f(x)=(3x-x^3)/2$. The map has 3 fixed points namely -1,0 & 1. The $x = y$ line cuts the function graph at the fixed points. The orbits with initial values $x = 1.6$ and $x = 1.8$ are drawn. The orbit with initial value 1.6 converges to the sink at 1. The orbit with initial value 1.8 converges to the sink at -1. The results are in accordance with the reference [5].

# Chapter - 3  The Family Of Logistic Maps

## (One Dimensional)

### 3.1. Logistic Map

The family of logistic map $g(x)=ax(1-x)$ is investigated for different 'a' values ranging from 0 to 4. The map is iterated 50 times for each 'a' value and tabulated. The calculations were performed by Microsoft Excel software. Each table is also given in graphical form.

**Table – 3   Logistic Map**

| n | a= 2.8 | a= 3.3 | a= 3.5 | a= 3.9 |
|---|--------|--------|--------|--------|
| n | x | x | x | x |
| 0 | 0.95 | 0.95 | 0.95 | 0.95 |
| 1 | 0.133 | 0.15675 | 0.16625 | 0.18525 |
| 2 | 0.322871 | 0.436192 | 0.485138 | 0.588637 |
| 3 | 0.612151 | 0.811564 | 0.874227 | 0.94436 |
| 4 | 0.664782 | 0.504661 | 0.38484 | 0.204923 |
| 5 | 0.623971 | 0.824928 | 0.828583 | 0.635424 |
| 6 | 0.656967 | 0.476591 | 0.497115 | 0.903475 |
| 7 | 0.631012 | 0.823192 | 0.874971 | 0.340111 |
| 8 | 0.651941 | 0.480305 | 0.382889 | 0.875299 |
| 9 | 0.635359 | 0.82372 | 0.826997 | 0.425688 |
| 10 | 0.648698 | 0.479178 | 0.500754 | 0.953463 |
| 11 | 0.638089 | 0.823569 | 0.874998 | 0.173047 |
| 12 | 0.646608 | 0.4795 | 0.382818 | 0.558097 |
| 13 | 0.639817 | 0.823613 | 0.826939 | 0.961836 |
| 14 | 0.645263 | 0.479406 | 0.500888 | 0.143158 |
| 15 | 0.640916 | 0.8236 | 0.874997 | 0.478388 |
| 16 | 0.6444 | 0.479433 | 0.38282 | 0.973178 |
| 17 | 0.641617 | 0.823604 | 0.826941 | 0.101799 |
| 18 | 0.643845 | 0.479425 | 0.500884 | 0.356599 |
| 19 | 0.642064 | 0.823603 | 0.874997 | 0.894801 |
| 20 | 0.64349 | 0.479428 | 0.38282 | 0.367115 |
| 21 | 0.64235 | 0.823603 | 0.826941 | 0.906132 |
| 22 | 0.643262 | 0.479427 | 0.500884 | 0.331721 |
| 23 | 0.642533 | 0.823603 | 0.874997 | 0.864561 |
| 24 | 0.643117 | 0.479427 | 0.38282 | 0.456672 |
| 25 | 0.642649 | 0.823603 | 0.826941 | 0.967678 |
| 26 | 0.643023 | 0.479427 | 0.500884 | 0.12198 |
| 27 | 0.642724 | 0.823603 | 0.874997 | 0.417693 |
| 28 | 0.642963 | 0.479427 | 0.38282 | 0.94858 |

| 29 | 0.642772 | 0.823603 | 0.826941 | 0.190228 |
| 30 | 0.642925 | 0.479427 | 0.500884 | 0.600761 |
| 31 | 0.642803 | 0.823603 | 0.874997 | 0.935404 |
| 32 | 0.642901 | 0.479427 | 0.38282  | 0.23565  |
| 33 | 0.642822 | 0.823603 | 0.826941 | 0.702464 |
| 34 | 0.642885 | 0.479427 | 0.500884 | 0.815132 |
| 35 | 0.642835 | 0.823603 | 0.874997 | 0.587698 |
| 36 | 0.642875 | 0.479427 | 0.38282  | 0.945005 |
| 37 | 0.642843 | 0.823603 | 0.826941 | 0.202685 |
| 38 | 0.642869 | 0.479427 | 0.500884 | 0.630254 |
| 39 | 0.642848 | 0.823603 | 0.874997 | 0.908832 |
| 40 | 0.642864 | 0.479427 | 0.38282  | 0.323138 |
| 41 | 0.642851 | 0.823603 | 0.826941 | 0.853008 |
| 42 | 0.642862 | 0.479427 | 0.500884 | 0.489003 |
| 43 | 0.642853 | 0.823603 | 0.874997 | 0.974528 |
| 44 | 0.64286  | 0.479427 | 0.38282  | 0.096809 |
| 45 | 0.642855 | 0.823603 | 0.826941 | 0.341005 |
| 46 | 0.642859 | 0.479427 | 0.500884 | 0.87641  |
| 47 | 0.642856 | 0.823603 | 0.874997 | 0.422431 |
| 48 | 0.642858 | 0.479427 | 0.38282  | 0.951534 |
| 49 | 0.642856 | 0.823603 | 0.826941 | 0.179858 |
| 50 | 0.642858 | 0.479427 | 0.500884 | 0.575285 |

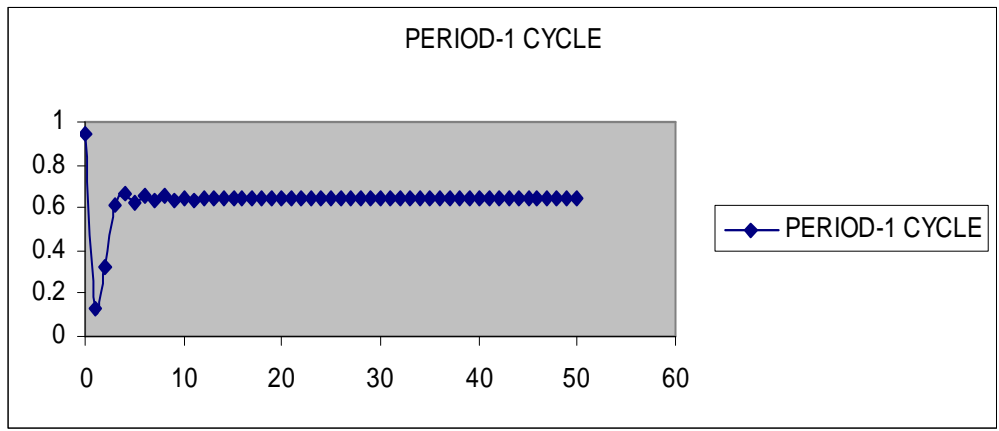

**Fig.5**

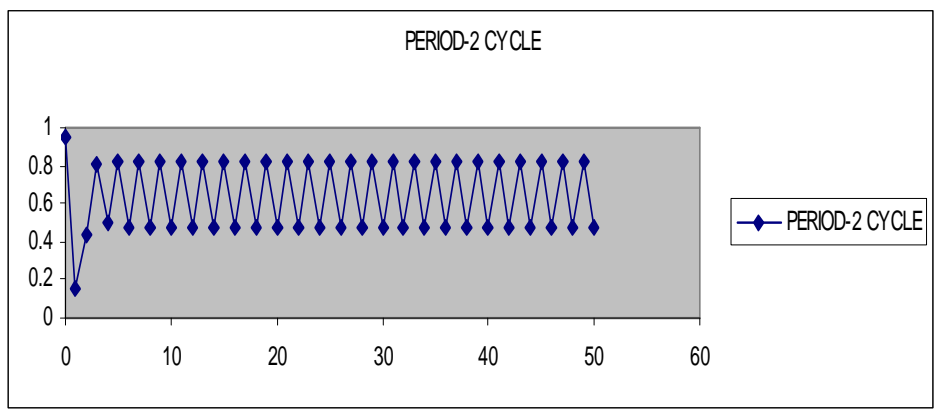

**Fig.6**

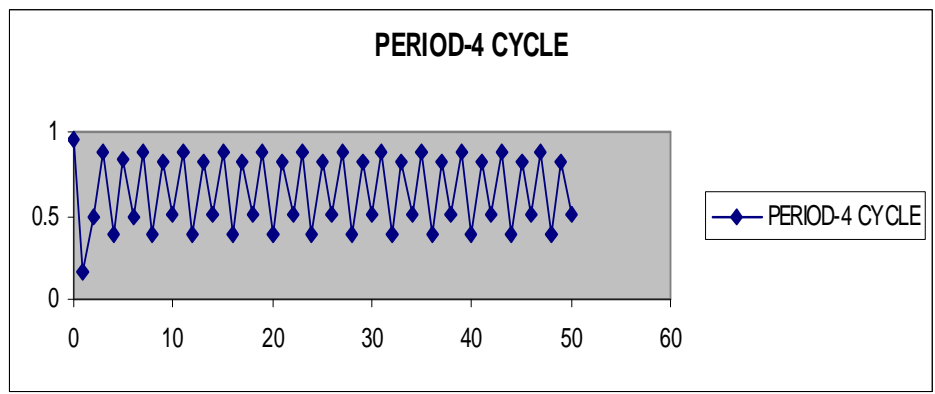

**Fig.7**

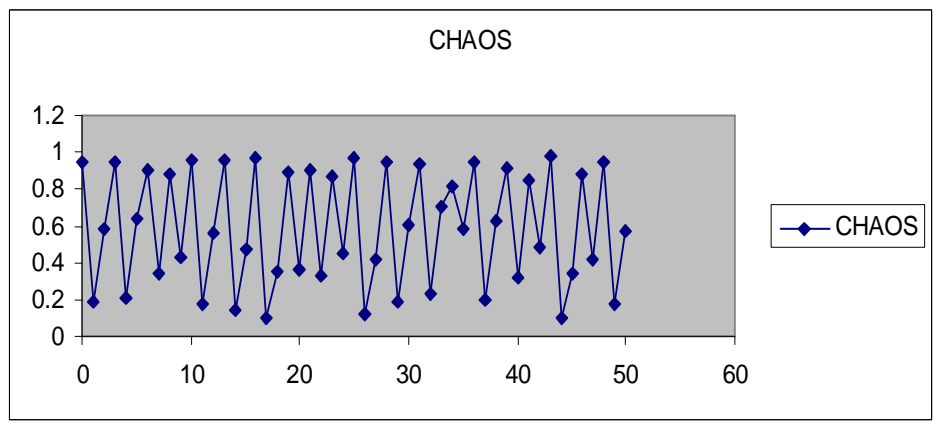

**Fig.8**

### 3.2. Bifurcation

The limiting behaviour of orbits for values of $a$ in the range $1 < a < 4$ is given by a diagram which is known as bifurcation diagram.

Steps to Produce Bifurcation Diagram

(1) Choose a value of a starting with a=1.
(2) Choose $x$ at random in [0,1]
(3) Calculate the orbit of $x$ under $g_a(x)$. (Table – 4)
(4) Ignore the first 100 iterates and plot the orbit beginning with iterate (0), Then $a$ is incremented and the same procedure is followed. The points that are plotted will approximate either fixed or periodic sinks or other attracting sets.

The Bifurcation diagram shows the birth, evolution and death of attracting sets. The $101^{st}$ iterate value for each '$a$' value is given in table - 5 to draw the bifurcation diagram. The calculations were performed by Microsoft Excel software.

**Table – 4   To Produce Bifurcation Diagram**

| a = 2 | a = 3 | a = 3.5 | a = 4 |
|---|---|---|---|
| 0.2 | 0.2 | 0.2 | 0.2 |
| 0.32 | 0.48 | 0.56 | 0.64 |
| 0.4352 | 0.7488 | 0.8624 | 0.9216 |
| 0.491602 | 0.564296 | 0.415332 | 0.289014 |
| 0.499859 | 0.737598 | 0.84991 | 0.821939 |
| 0.5 | 0.580641 | 0.446472 | 0.585421 |
| 0.5 | 0.730491 | 0.864971 | 0.970813 |
| 0.5 | 0.590622 | 0.408785 | 0.113339 |
| 0.5 | 0.725363 | 0.84588 | 0.401974 |
| 0.5 | 0.597634 | 0.456285 | 0.961563 |
| 0.5 | 0.721403 | 0.868312 | 0.147837 |
| 0.5 | 0.602943 | 0.400213 | 0.503924 |
| 0.5 | 0.718208 | 0.840149 | 0.999938 |
| 0.5 | 0.607155 | 0.470046 | 0.000246 |
| 0.5 | 0.715553 | 0.87186 | 0.000985 |
| 0.5 | 0.61061 | 0.391022 | 0.003936 |
| 0.5 | 0.713296 | 0.833433 | 0.015682 |
| 0.5 | 0.613514 | 0.485879 | 0.061745 |
| 0.5 | 0.711343 | 0.874302 | 0.23173 |
| 0.5 | 0.616002 | 0.384643 | 0.712124 |
| 0.5 | 0.709631 | 0.828425 | 0.820014 |
| 0.5 | 0.618165 | 0.49748 | 0.590364 |
| 0.5 | 0.708111 | 0.874978 | 0.967337 |
| 0.5 | 0.620069 | 0.382871 | 0.126384 |
| 0.5 | 0.70675 | 0.826983 | 0.441645 |
| 0.5 | 0.621763 | 0.500788 | 0.986379 |
| 0.5 | 0.705521 | 0.874998 | 0.053742 |
| 0.5 | 0.623283 | 0.382818 | 0.203415 |
| 0.5 | 0.704404 | 0.826939 | 0.64815 |
| 0.5 | 0.624657 | 0.500887 | 0.912207 |
| 0.5 | 0.703382 | 0.874997 | 0.320342 |
| 0.5 | 0.625908 | 0.38282 | 0.870893 |
| 0.5 | 0.702442 | 0.826941 | 0.449755 |
| 0.5 | 0.627052 | 0.500884 | 0.989902 |
| 0.5 | 0.701573 | 0.874997 | 0.039986 |
| 0.5 | 0.628104 | 0.38282 | 0.153547 |
| 0.5 | 0.700768 | 0.826941 | 0.519882 |
| 0.5 | 0.629077 | 0.500884 | 0.998419 |
| 0.5 | 0.700017 | 0.874997 | 0.006315 |
| 0.5 | 0.629979 | 0.38282 | 0.025099 |
| 0.5 | 0.699316 | 0.826941 | 0.097874 |
| 0.5 | 0.630819 | 0.500884 | 0.35318 |
| 0.5 | 0.698659 | 0.874997 | 0.913776 |
| 0.5 | 0.631604 | 0.38282 | 0.315159 |
| 0.5 | 0.698041 | 0.826941 | 0.863335 |
| 0.5 | 0.632339 | 0.500884 | 0.47195 |
| 0.5 | 0.697459 | 0.874997 | 0.996853 |
| 0.5 | 0.633029 | 0.38282 | 0.01255 |
| 0.5 | 0.69691 | 0.826941 | 0.049568 |
| 0.5 | 0.63368 | 0.500884 | 0.188445 |
| 0.5 | 0.696389 | 0.874997 | 0.611733 |
| 0.5 | 0.634294 | 0.38282 | 0.950063 |
| 0.5 | 0.695895 | 0.826941 | 0.189772 |
| 0.5 | 0.634875 | 0.500884 | 0.615035 |
| 0.5 | 0.695426 | 0.874997 | 0.947067 |
| 0.5 | 0.635426 | 0.38282 | 0.200523 |
| 0.5 | 0.69498 | 0.826941 | 0.641254 |
| 0.5 | 0.635949 | 0.500884 | 0.920189 |
| 0.5 | 0.694554 | 0.874997 | 0.293764 |
| 0.5 | 0.636447 | 0.38282 | 0.829868 |
| 0.5 | 0.694147 | 0.826941 | 0.56475 |
| 0.5 | 0.636921 | 0.500884 | 0.98323 |
| 0.5 | 0.693758 | 0.874997 | 0.065955 |
| 0.5 | 0.637373 | 0.38282 | 0.246421 |
| 0.5 | 0.693386 | 0.826941 | 0.742791 |
| 0.5 | 0.637806 | 0.500884 | 0.76421 |
| 0.5 | 0.693028 | 0.874997 | 0.720773 |
| 0.5 | 0.63822 | 0.38282 | 0.805037 |
| 0.5 | 0.692686 | 0.826941 | 0.62781 |
| 0.5 | 0.638617 | 0.500884 | 0.934659 |
| 0.5 | 0.692356 | 0.874997 | 0.244287 |
| 0.5 | 0.638997 | 0.38282 | 0.738444 |
| 0.5 | 0.692039 | 0.826941 | 0.772578 |
| 0.5 | 0.639363 | 0.500884 | 0.702804 |
| 0.5 | 0.691734 | 0.874997 | 0.835482 |
| 0.5 | 0.639714 | 0.38282 | 0.549808 |
| 0.5 | 0.69144 | 0.826941 | 0.990077 |
| 0.5 | 0.640052 | 0.500884 | 0.0393 |
| 0.5 | 0.691156 | 0.874997 | 0.151022 |
| 0.5 | 0.640378 | 0.38282 | 0.512857 |
| 0.5 | 0.690882 | 0.826941 | 0.999339 |
| 0.5 | 0.640692 | 0.500884 | 0.002643 |
| 0.5 | 0.690617 | 0.874997 | 0.010545 |
| 0.5 | 0.640995 | 0.38282 | 0.041733 |
| 0.5 | 0.690361 | 0.826941 | 0.159967 |
| 0.5 | 0.641288 | 0.500884 | 0.53751 |
| 0.5 | 0.690113 | 0.874997 | 0.994372 |
| 0.5 | 0.641571 | 0.38282 | 0.022386 |
| 0.5 | 0.689873 | 0.826941 | 0.087538 |
| 0.5 | 0.641845 | 0.500884 | 0.319501 |
| 0.5 | 0.68964 | 0.874997 | 0.869681 |
| 0.5 | 0.64211 | 0.38282 | 0.453344 |
| 0.5 | 0.689414 | 0.826941 | 0.991293 |
| 0.5 | 0.642367 | 0.500884 | 0.034525 |
| 0.5 | 0.689195 | 0.874997 | 0.13333 |
| 0.5 | 0.642616 | 0.38282 | 0.462213 |
| 0.5 | 0.688982 | 0.826941 | 0.994289 |
| 0.5 | 0.642857 | 0.500884 | 0.022715 |
| 0.5 | 0.688776 | 0.874997 | 0.088795 |
| 0.5 | 0.643091 | 0.38282 | 0.323642 |
| 0.5 | 0.688575 | 0.826941 | 0.875591 |
| 0.5 | 0.643319 | 0.500884 | 0.435725 |
| 0.5 | 0.688379 | 0.874997 | 0.983475 |
| 0.5 | 0.64354 | 0.38282 | 0.065008 |
| 0.5 | 0.688189 | 0.826941 | 0.243127 |
| 0.5 | 0.643755 | 0.500884 | 0.736064 |

## Table – 5   101$^{st}$ Iterate Value

| a   | x    |
|-----|------|
| 2   | 0.5  |
| 3   | 0.69 |
| 3   | 0.64 |
| 3.5 | 0.83 |
| 3.5 | 0.5  |
| 3.5 | 0.87 |
| 3.5 | 0.38 |
| 4   | 0.88 |
| 4   | 0.44 |
| 4   | 0.98 |
| 4   | 0.07 |
| 4   | 0.24 |
| 4   | 0.74 |
| 4   | 0.77 |
| 4   | 0.69 |

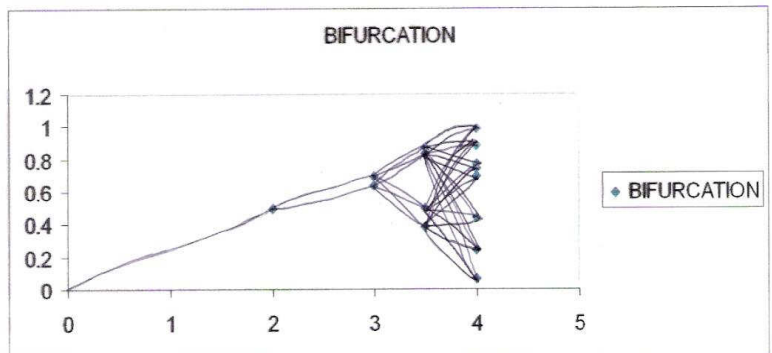

**Fig.9**

## Results and Discussion

When $0 < a < 1$, the map has a sink at $x=0$ and every initial condition between 0 and 1 is attracted to this sink. In other words with small reproduction rates, small populations tend to die out. If $1 < a < 3$, the map has a sink at $x=(a-1)/a$. Small proportions grow to a steady state of $x=(a-1)/a$ for $a>3$, the fixed point is unstable and a period two sink takes its place for a=3.3. When a grows above $1+\sqrt{6} \approx 3.45$, the period two sink also becomes unstable. Many new periodic orbits come into existence as $a$ is increased from 3.45 to 4.

## Chapter 4 – Two Dimensional Maps

Much of the Chaotic Phenomena present in differential equations can be approached through reduction by time – T maps and poincare maps. Poincare maps of differential equations can be found as well in a two dimensional quadratic map which is much easier to simulate on a computer. One such map is the Henon map which is given as $f(x,y)=(a-x^2+by,x)$.

The map has two inputs $x,y$ and two outputs the new $x,y$. The new $y$ is just the old $x$ but the new $x$ is a non linear function of the old $x$ and $y$. The letters $a$ and $b$ represent parameters that are held fixed as the map is iterated.

Henon's remarkable discovery is "barely nonlinear" map. It displays an impressive breadth of complex phenomena. In its way, the Henon map is to two dimensional dynamics what the logistic map $G(x)=4x(1-x)$ is to one dimensional dynamics and it continues to be a catalyst for deeper understanding of nonlinear systems.

**Sinks, Sources and Saddles**

Along with the sink and source a new type of fixed point is there, which cannot occur in a one dimensional state space. This type of fixed point which is called as a saddle has at least one attracting direction and at least one repelling direction.

**Definition for Saddle:**

Let A be a linear map. A is hyperbolic if A has no eigen values of absolute value one. If a hyperbolic map A has at least one eigen value of absolute value greater than one and at least one eigen value of absolute value smaller than one, then the origin is called a saddle.

There are three types of hyperbolic maps.

One for which the origin is a sink
One for which the origin is a source
3. One for which the origin is a saddle

Hyperbolic linear maps are important objects of study because they have well defined expanding and contracting directions.

**4.1. Henon Map**

*It is given by $f_{a,b} = (a-x^2+by, x)$*

*Assume $a = 0$; $b = 0.4$;*

to find the fixed points

$-x^2 + by = x$

$-x^2 + 0.4x = x$

$x^2 + 0.6x = 0$

$x(x+0.6) = 0$

$x = 0 ;\ x = -0.6;$

Therefore, the fixed points are

(0, 0) and (-0.6, -0.6).

to check for the fixed point (0, 0):

The Jacobian matrix is

$$Df(x,y) = \begin{pmatrix} -2x & b \\ 1 & 0 \end{pmatrix}$$

$$Df(0, 0) = \begin{pmatrix} 0 & 0.4 \\ 1 & 0 \end{pmatrix}$$

$$\begin{pmatrix} -\lambda & 0.4 \\ 1 & 0-\lambda \end{pmatrix}$$

$\lambda^2 - 0.4 = 0$

$\lambda^2 = 0.4$

$\lambda = \pm \sqrt{0.4} = \pm 0.632$

The eigen values are +0.632 and -0.632. Both are less than 1.

So the fixed point (0, 0) is a sink.

*to check for the fixed point ( -0.6, -0.6)*

The Jacobian matrix is

$$Df(x,y) = \begin{pmatrix} -2x & b \\ 1 & 0 \end{pmatrix}$$

$$Df(-0.6, -0.6) = \begin{pmatrix} 1.2 & 0.4 \\ 1 & 0 \end{pmatrix}$$

$$\begin{pmatrix} 1.2-\lambda & 0.4 \\ 1 & -\lambda \end{pmatrix} = -1.2\lambda + \lambda^2 = 0.4$$

$$\lambda^2 - 1.2\lambda - 0.4 = 0$$

$$\lambda = \frac{1.2 \pm \sqrt{1.44 + 1.6}}{2}$$

eigen values = 1.472, -0.271

One is greater than one and other is <1.

So, the fixed point (-0.6, -0.6) is a saddle.

For b=0.4 and a>0.85 the attractors of the Henon map become more complex [Table – 6]. When the period - two orbit becomes unstable, it is replaced with an attracting period 4 orbit, then a period eight orbit etc.

## Table – 6  Henon Map

| Period 1 | | Period 2 | | Period 4 | |
|---|---|---|---|---|---|
| a=1.2 | b= -0.3 | a=1.28 | b= -0.3 | a=0.9 | b=0.4 |
| x | y | x | y | x | y |
| 0.623774 | 0.623774 | 0.532437 | 0.766773 | -0.23842 | 0.932101 |
| 0.623774 | 0.623774 | 0.766478 | 0.532437 | 1.215997 | -0.23842 |
| 0.623774 | 0.623774 | 0.53278 | 0.766478 | -0.67402 | 1.215997 |
| 0.623774 | 0.623774 | 0.766202 | 0.53278 | 0.932101 | -0.67402 |
| 0.623774 | 0.623774 | 0.5331 | 0.766202 | -0.23842 | 0.932101 |
| 0.623774 | 0.623774 | 0.765944 | 0.5331 | 1.215997 | -0.23842 |
| 0.623774 | 0.623774 | 0.5334 | 0.765944 | -0.67402 | 1.215997 |
| 0.623774 | 0.623774 | 0.765701 | 0.5334 | 0.932101 | -0.67402 |
| | | | | -0.23842 | 0.932101 |
| | | | | 1.215997 | -0.23842 |
| | | | | -0.67402 | 1.215997 |
| | | | | 0.932101 | -0.67402 |
| | | | | | |
| Period 10 | | Period 16 | | | |
| a=1.0293 | b=0.4 | a=0.988 | b=0.4 | | |
| x | y | x | y | | |
| -0.75447 | 1.252643 | -0.02265 | 0.82423 | | |
| 0.961137 | -0.75447 | 1.317179 | -0.02265 | | |
| -0.19627 | 0.961137 | -0.75602 | 1.317179 | | |
| 1.375232 | -0.19627 | 0.943303 | -0.75602 | | |
| -0.94047 | 1.375232 | -0.20423 | 0.943303 | | |
| 0.694904 | -0.94047 | 1.323612 | -0.20423 | | |
| 0.170219 | 0.694904 | -0.84564 | 1.323612 | | |
| 1.278287 | 0.170219 | 0.802338 | -0.84564 | | |
| -0.53663 | 1.278287 | 0.005997 | 0.802338 | | |
| 1.252643 | -0.53663 | 1.308899 | 0.005997 | | |
| -0.75447 | 1.252643 | -0.72282 | 1.308899 | | |
| 0.961138 | -0.75447 | 0.989093 | -0.72282 | | |
| -0.19627 | 0.961138 | -0.27943 | 0.989093 | | |
| 1.375232 | -0.19627 | 1.035555 | -0.27943 | | |
| -0.94047 | 1.375232 | -0.82825 | 1.035555 | | |
| 0.694904 | -0.94047 | 0.82423 | -0.82825 | | |
| 0.170219 | 0.694904 | -0.02265 | 0.82423 | | |
| 1.278287 | 0.170219 | 1.317179 | -0.02265 | | |
| -0.53663 | 1.278287 | -0.75602 | 1.317179 | | |
| 1.252643 | -0.53663 | 0.943302 | -0.75602 | | |
| -0.75447 | 1.252643 | | | | |
| 0.961138 | -0.75447 | | | | |
| -0.19627 | 0.961138 | | | | |
| 1.375232 | -0.19627 | | | | |

**Results and Discussion**

After 500 iterations the Henon map displays a single attracting orbit for a particular value of the parameter 'a'. From the Table – 6, we see that for b = 0.4 and a> 0.85, the attractors of the Henon map become more complex. When a = 0.9, there is a period 4 sink. When a = 0.988 there is an attracting period 16 sink and when a = 1.0293, there is a period 10 sink. Thus, the periodic points are the key to many of the properties of a map [5].

### 4.2. Simple Dynamical Models

In this chapter, we modeled two physical processes with maps. One of the most important uses of maps in scientific applications is to assist in the study of a differential equation model. A map describes the time evolution of a system by expressing its state as a function of its previous state. Instead of expressing the current state as a function of the previous state, a differential equation expresses the rate of change of the current state as a function of the current state.

A simple illustration of this type of dependence is Newton's law of cooling. Consider the state $x$ consisting of the difference between the temperature of a warm object and the temperature of its surroundings. The rate of change of this temperature difference is negatively proportional to the temperature difference itself:

$x = ax$   where a<0.

The solution of this equation is $x(t) = x(0)e^{at}$, meaning that the temperature difference $x$ decays exponentially in time. This is a linear differential equation, since the terms involving the state $x$ and its derivatives are linear terms. Since it is a linear differential equation, whatever be the initial condition $x(0)$, there are no attracting fixed points.

Another familiar example, which yields a nonlinear differential equation, is that of a pendulum. The pendulum bob hangs from a pivot, which constrains it to move along a circle. The acceleration of the pendulum bob in the tangential direction is proportional to the component of the gravitational downward force in the tangential direction, which in turn depends on the current position of the pendulum. This relation of the second derivative of the angular position with the angular position itself is one of the most fundamental equations in science. The pendulum is an example of a nonlinear oscillator. Other nonlinear oscillators that satisfy the same general type of differential equation include electric circuits and feedback systems. Newton's law of motion F=ma is used to find the pendulum equation. If l is the length of the pendulum, and θ is the angle of the pendulum, and m is the mass of the pendulum, then the component of acceleration tangent to the circle is lθ̈, because the component of position tangent to the circle is lθ. The component of force along the direction of motion is mgsinθ. It is a restoring force, meaning that it is directed in the opposite direction from the displacement of the variable θ. If the first and second time derivatives of θ are θ̇ and θ̈, the differential equation governing the frictionless pendulum is mlθ̈ = -mg sinθ, according to Newton's law of motion. To simplify, we will use a pendulum of length l=g. Now the equation becomes θ̈ = - sin θ . The Table -7 gives the solution.

**Table – 7   Undamped Simple Pendulam**

| θ | θ̇ |
|---|---|
| 0.3 | 0.7 |
| 0.29552 | 0.755336 |
| 0.291238 | 0.756651 |
| 0.287138 | 0.757889 |
| 0.283208 | 0.759058 |
| 0.279438 | 0.760164 |
| 0.275815 | 0.761211 |
| 0.272331 | 0.762204 |
| 0.268978 | 0.763146 |
| 0.265746 | 0.764043 |

Again, there are no fixed points for this nonlinear, two dimensional map.

Now, we add the damping term -c $\dot{\theta}$, corresponding to friction at the pivot, and a periodic term r sin t which is an external force constantly providing energy to the pendulum. The resulting equation, which we call the forced damped pendulum model, is

$$\ddot{\theta} = -c\dot{\theta} - \sin\theta + r \sin t$$

An approximate solution for this equation is found for c =0.2 and r =1.66. With the initial condition $\theta$ = 0.3 and $\dot{\theta}$ = 0.7, we find from the Table -8 that periodic orbits exist.

Table – 8   <u>Damped Simple Pendulam</u>

| $\theta$ | $\dot{\theta}$ |
|---|---|
| 1.136717 | -0.28691 |
| 0.124961 | 0.220575 |
| -1.631522 | 0.792203 |
| 1.097059 | -0.260688 |
| 0.034552 | 0.256215 |
| -1.657816 | 0.799403 |
| 1.136717 | -0.28691 |
| 0.124961 | 0.220575 |
| -1.631522 | 0.792203 |
| 1.097059 | -0.260688 |
| 0.034552 | 0.256215 |
| -1.657816 | 0.799403 |
| 1.136717 | -0.28691 |
| 0.124961 | 0.220575 |
| -1.631522 | 0.792203 |
| 1.097059 | -0.260688 |
| 0.034552 | 0.256215 |
| -1.657816 | 0.799403 |
| 1.136717 | -0.28691 |
| 0.124961 | 0.220575 |
| -1.631522 | 0.792203 |
| 1.097059 | -0.260688 |
| 0.034552 | 0.256215 |
| -1.657816 | 0.799403 |

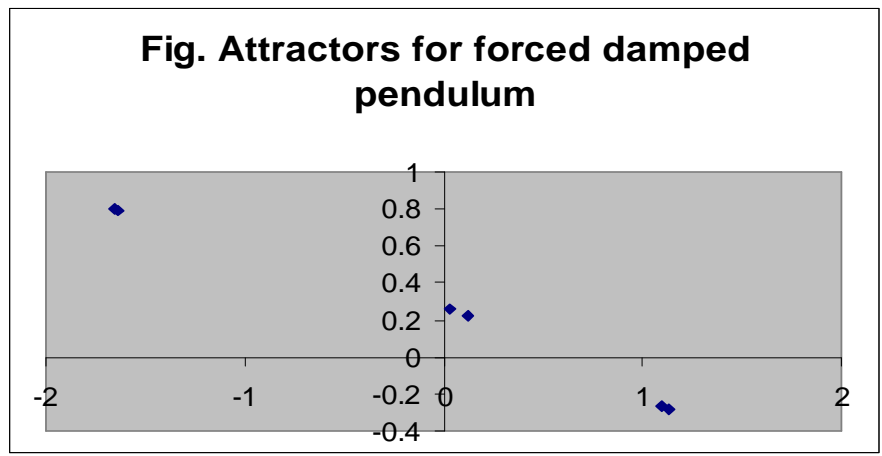

**Fig.10**

It is clear from the above discussion that, period multiplication occurs for forced damped simple pendulum which leads to chaotic solution.

The exact solution can be obtained by solving the differential equation by using Runge-kutta method.

Thus an analysis is done on different dynamical systems and the condition under which bifurcation and chaos occur in such systems.

# Appendix

## Program For Henon Map (In C++)

```cpp
#include<iostream.h>
#include<conio.h>
void main( )
{
int n,i;
float x=0;
float y=0;
float a=0.9;
float b=0.4;
float nx,ny;
clrscr( );
cout<< "enter i " ;
cin>>i ;
cout<<"\n";
for(n=1;n<=i;n++)
{
nx=a-((x)*(x))+(b*y);
ny=x;
cout<<x<<"\t\t\t"<<y<<"\n";
x=nx;
y=ny
}
getch( );
}
```

* * *   *   * * *